\documentclass[aps,pre,twocolumn,showpacs,superscriptaddress,groupedaddress]{revtex4}  % for review and submission

\usepackage[T2A]{fontenc}
\usepackage[cp1251]{inputenc}
\usepackage[english]{babel}
\usepackage{graphicx}
\usepackage{subfigure}
\usepackage{multirow}
\usepackage{amsmath}
\usepackage{amssymb}
\begin{document}
\title{Sedimentation stability and aging of aqueous dispersions of Laponite in the presence of cetyltrimethylammonium bromide}

\author{V. Savenko}
%\email{savenkovolod@mail.ru}
\affiliation{Taras Shevchenko Kiev National University, Department of Physics, 2, av. Academician Glushkov, Kyiv 031127, Ukraine}

\author{L. Bulavin}
%\email{bulavin221@gmail.com }
\affiliation{Taras Shevchenko Kiev National University, Department of Physics, 2, av. Academician Glushkov, Kyiv 031127, Ukraine}

\author{M. Rawiso}
%\email{michel.rawiso@ics-cnrs.unistra.fr}
\affiliation{Institut Charles Sadron, UPR22-CNRS,  Universite de Strasbourg, 23 rue du Loess, BP 84047-67034 Strasbourg Cedex 2, France}

\author{M. Loginov}
%\email{maksym.loginov@gmail.com}
\affiliation{Institute of Biocolloidal Chemistry named after F.D. Ovcharenko, NAS of Ukraine, 42, blvr. Vernadskogo, Kyiv 03142, Ukraine}
\affiliation{Departement de Genie Chimique, Universite de Technologie de Compiegne, Centre de Recherche de Royallieu, B.P. 20529-60205 Compiegne Cedex, France}

\author{E. Vorobiev}
%\email{eugene.vorobiev@utc.fr}
\affiliation{Departement de Genie Chimique, Universite de Technologie de Compiegne, Centre de Recherche de Royallieu, B.P. 20529-60205 Compiegne Cedex, France}

\author{N. I. Lebovka}
\email[Corresponding author: ]{lebovka@gmail.com}
%\affiliation{Taras Shevchenko Kiev National University, Department of Physics, 2, av. Academician Glushkov, Kyiv 031127, Ukraine}
\affiliation{Institute of Biocolloidal Chemistry named after F.D. Ovcharenko, NAS of Ukraine, 42, blvr. Vernadskogo, Kyiv 03142, Ukraine}
\affiliation{Departement de Genie Chimique, Universite de Technologie de Compiegne, Centre de Recherche de Royallieu, B.P. 20529-60205 Compiegne Cedex, France}
\date{\today}

\begin{abstract}

This work discusses the sedimentation stability and aging of aqueous suspension of Laponite in the presence of cetyltrimethylammonium bromide (CTAB). The concentration of Laponite was fixed at the constant level $C_l=2$ \%wt, which corresponds to the threshold between equilibrium gel IG$_1$ and repulsive gel IG$_2$ phases. The concentration of CTAB $C_s$ was within 0-0.3 \%wt. In the presence of CTAB the Laponite aqueous suspensions were unstable against sedimentation and they separated out into upper and bottom layers (U- and B-layers, respectively). The dynamic light scattering technique revealed that the addition of CTAB even at rather small concentration, $C_s=0.0164$ \%wt ($0.03 CEC$), induced noticeable changes in the aging dynamics of U-layer, and it was explained by equilibration of CTAB molecules that were initially non-uniformly distributed between different Laponite particles.  Accelerated stability analysis by means of analytical centrifugation with rotor speed  $\omega=500-4000$ rpm revealed three sedimentation regimes:  continuous (I, $C_s<0.14$ \%wt), zone-like (II, $0.14<C_s<0.2$ \%wt) and gel-like (III, $C_s >0.2$ \%wt). It was demonstrated that B-layer was "soft" in the zone-like regime. The increase of  $\omega$ resulted in its supplementary compressing and the collapse of "soft" sediment above certain critical centrifugal acceleration.
\end{abstract}

%\pacs{64.60.ah, 64.60.De, 68.35.Rh, 61.43.Bn}
\keywords{CTAB; Laponite; aqueous suspensions; aging; sedimentation stability; dynamic light scattering; analytical centrifugation}

\maketitle

\section{\label{sec:introduction}Introduction}

Nowadays the aqueous suspensions of Laponite disks attract great fundamental interest as model colloidal systems with complex aging dynamic and nonergodic arrested states \cite{Kegel2011, Ruzicka2011, Ruzicka2011a}. Phase diagrams of these suspensions were extensively studied, discussed and revisited in recent years \cite{Mourchid1995, Mourchid1998, Bonn1999, Levitz2000, Ruzicka2004, Ruzicka2004a, Mongondry2005,   Ruzicka2006, Cummins2007, Jabbari-Farouji2008, Kegel2011, Ruzicka2011, Ruzicka2011a}.  They are commonly specified in terms of three main variables: concentration of Laponite, $C_l$, concentration of ionic component, $C_s$, and time of aging, $t_a$. Depending on $C_l$, the following main phase states were observed in pure Laponite suspensions (i.e., at $C_s=0$) with long aging time (months, years) \cite{Kegel2011, Ruzicka2011, Ruzicka2011a}:
\begin{itemize}
  \item mixed state of separated sol and gel  phases ($C_l<1$ \%wt),
  \item stable gel, equilibrium gel IG$_1$ (1  \%wt $<C_l< 2$ \%wt),
  \item Wigner glass or repulsive gel IG$_2$ (2 \%wt $< C_l< 3$ \%wt),
   \item nematic gel (>3 \%wt).
\end{itemize}

Addition of ionic components to Laponite suspensions intensify their ageing and results in a decrease in the time of transition into the arrested state. E.g., addition of NaCl up to the level above 20 mM accelerates aggregation and sedimentation processes and makes suspensions turbid even at low Laponite concentrations \cite{Nicolai2000, Nicolai2001}. The time of transition into the arrested state also directly depends on $C_l$ and $C_s$. Very interesting is the phase behaviour of Laponite suspensions in the presence of surfactant ions
that lead to an organic modification of clay particles \cite{Hanley1997, Patil2008}.

The alkylammonium salts are widely used as organic modifier since they have high ability to adsorb on the Laponite surface \cite{Liu2010, Zhang2008}. Previous studies indicated that introduction of cetyltrimethylammonium bromide (CTAB) in aqueous Laponite suspensions caused fast aggregation of Laponite particles and violate sedimentation stability of suspensions. At certain threshold concentration of CTAB ( $\approx 0.3 CEC$, where $CEC$ is the cation exchange capacity of Laponite), separation of CTAB+Laponite suspensions into clear upper (U) and turbid bottom (B) layers was observed \cite{Zhang2008}. In this work, the 1-week aged Laponite suspensions were diluted by CTAB solutions. However, the Laponite suspensions are unstable against aging \cite{Abou2001, Guillermic2009, Joshi2008, Knaebel2000, Labanda2008, Martin2012, Shahin2012}, so, the aging processes may influence the phase behaviour and  sedimentation separation of CTAB+Laponite suspensions.

In this paper, the detailed study of the sedimentation stability and ageing of aqueous suspensions of disk-like Laponite particles in the presence of CTAB surfactant was done. The concentration of Laponite was fixed at a rather high level of $C_l=2$ \%wt that roughly corresponded to the the boundary between isotropic or equilibrium gel IG$_1$ and repulsive or Wigner glass gel IG$_2$ phases. The concentration of CTAB $C_s$ was $< 0.3$ \%wt ( $\approx 0.55CEC$).  The dynamic light scattering technique revealed that introduction of CTAB even in a rather small concentration, $C_s=0.0164$ \%wt, induced the noticeable changes in the aging dynamics of Laponite suspensions. The express analysis of sedimentation stability was done using analytical centrifugation technique with the rotor speed $\omega$   varied within 500-4000 rpm. Different sedimentation regimes that were dependent on CTAB concentration $C_s$ were revealed and discussed.

\section{\label{sec:MM}Materials and methods}

\subsection{\label{sec:Mat}Materials}
The formula of CTAB (Fluka, Germany, 99.5 \%wt) is C$_{16}$H$_{33}$N(CH$_3$)$_3$Br. The molar mass of CTAB is $364.45$ g/mol, the length of its molecule is $2.33$ nm, and its critical micelle concentration (CMC) in water is  $0.04$ \%wt \cite{Zhang2008}.

The formula of Laponite RD (Rockwood Additives Ltd., UK) is Na$^+_{0.7}$[(Si${_8}$Mg$_{5.5}$Li$_{0.4}$O$_{20}$(OH)$_4$]$_{0.7}^-$. It is composed of charged disk-like sheets with thickness about $h_l = 1$ nm and average diameter about $d_l = 25 - 30$ nm \cite{Laponite1990}. The crystal structure of Laponite consists of octahedrally coordinated magnesium oxides sandwiched between two sheets of tetrahedrally coordinated silica. According to literature data, specific surface area $S_l$, determined by adsorption of methylene blue, and density  $\rho_l$ of Laponite were equal to $370$ m$^2$/g and $2530$ kg/m$^3$, respectively. In aqueous suspensions, the Laponite particles are charged highly heterogeneously. Their faces have constant negative charge, while the positive charge of their edges is pH-dependent. The negative surface charge of Laponite RD, defined as cation exchange capacity (CEC), is equal to $0.75$ meq/g. The positive charges are generally screened by the diffuse part of the double electric layer of Laponite faces. Their values decrease with a pH increase and they are neutralized at pH$>11$ \cite{Martin2002, Tawari2001}.

\begin{figure} [htbp]%Figure 1
\centering
\includegraphics[width=0.9 \linewidth]{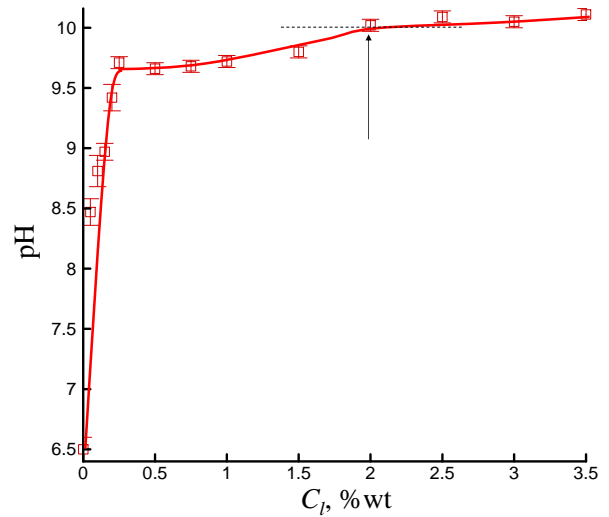}
\caption {\label{fig:Fig1}(Color online)
pH versus Laponite concentration, $C_l$, in the aqueous Laponite suspensions.}
\end{figure}

The Laponite powder was used for preparation of suspensions. It was dissolved in deionized ultrapure water (MilliQ, conductivity of 18.2  $\mu$S/cm) and then vigorously stirred during, at least, 20-30 minutes until reaching a homogeneous and transparent suspension. The pH of suspensions increased with increase of Laponite concentration (Fig. \ref{fig:Fig1}), however, it stabilized at the level of pH$\approx 10$ for concentrations $C_l$ above 2 \%wt. It is known that at smaller pH values Laponite particles may undergo exposure to dissolution, caused by CO$_2$, dissolved in the sample, that gets it from the ambient air \cite{Mourchid1998, Thompson1992}.  However, the high level of pH$ \approx 10$ that was used in this work allowed to expect that the particles were stable during the experiments \cite{Bonn1999}.

To prepare the CTAB+Laponite suspensions, CTAB and Laponite solutions in deionized ultrapure water (MilliQ) were mixed and vigorously stirred during, at least, 30 minutes.  The final concentration of Laponite, $C_l$, was fixed at 2 \%wt and concentration of CTAB, $C_s$, was varied within 0-0.3 \%wt. The value of $C_s$ ( \%wt) is related with CTAB concentration in suspension, expressed as a multiple of the clay $CEC$ (0.75 mmol/g) $C_s(CEC)$ and concentration of Laponite $C_l$( \%wt), through the following equation
\begin{equation}\label{Eq1}
C_s (CEC)= 3.663C_s/C_l,
\end{equation}
i.e., $C_s (CEC) \approx 1.83C_s$ at $C_l=2$ \%wt.

\subsection{\label{sec:Instrumentation}Instrumentation}

\subsubsection{\label{sec:DLS}Dynamic light scattering}

Dynamic light scattering (DLS) experiments were performed using ALV-5000 digital autocorrelator and HeNe laser with wavelength  $\lambda=633$ nm.  The scattered intensity $I(q, t_d)$ was measured as a function of decay time $t_d$ (1 $\mu s$ -10 s) and scattering vector $q=(4\pi n/\lambda)\sin(\theta /2)$, where $n$ is the refractive index of the solvent and  $\theta=90°$ is the scattering angle. The second-order autocorrelation function $g_2(t_d)$ was defined as

\begin{equation}\label{Eq2}
g_2(t_d)-1=<I(t_d)I(0)>/<I(0)^2>.
\end{equation}

In order to remove the effect of large aggregates, the suspensions were preliminary filtered (0.45 $ \mu m$, Millipore Millex AA) \cite{Bonn1999}.

\subsubsection{\label{sec:AC}Analytical centrifugation}

The sedimentation stability was investigated using analytical photocentrifuge LUMiSizer 610.0-135 (L.U.M. GmbH, Germany) that consisted of a centrifugal rotor with 12 optical cells, a light source (pulsed near-infrared light-emitting 880 nm diode and a light sensor). The operating principle of the analytical photocentrifuge is based on the measurement of profiles of light transmission $T(r)$ though the cell with the studied sample \cite{Lerche2007}. The value of $T(r)$ was measured continuously at various radial positions of the sample $r$. Mean light transmission, averaged over the height of the sample  $<T>$ , and volume-weighted cumulative distribution function $F$ of the particle size $d$ were calculated using a SepView 5.1 software (L.U.M. GmbH). Evolution of transmission profiles $T(r)$ and increase of mean light transmission  $<T>$  through the cell reflect continuous clarification of suspension, caused by settling of the Laponite aggregates in the centrifugal field.

The aqueous suspensions, weighting 0.4 g, were subjected to centrifugation in the rectangular polycarbonate optical cells, supplied by the photocentrifuge manufacturer. The optical path length was 2 mm and cross-sectional area was $1.6^. 10^{-5}$ m$^2$. The initial height of the sample in the cell was equal to $2.27^. 10^{-2}$ m. The radial distance from the axis of rotation to the centrifugal cell bottom $R$ was 130 mm. Centrifugation experiments were carried out at different rotor speeds  $\omega=500-4000$ rpm. The centrifugal acceleration at the bottom of the cell may be calculated as
\begin{equation}\label{Eq3}
g\approx 1.45^.10^{-4}\omega^2 g_o,
\end{equation}
where $g_o = 9.807$ m/s$^2$ is the gravity of the Earth.

\subsubsection{\label{sec:SA}Statistical analysis}
All the experiments were repeated, at least, three times. The Table Curve 2D software (Jandel Scientific, San Rafael, CA) was used to smooth the data and to determine their standard deviations. Mean and standard deviations were presented in the figures as error bars.

\section{\label{sec:RD}Results and discussion}
\subsection{\label{sec:Agening}Sedimentation stability and aging in Earth gravity}

\begin{figure} [htbp]%Figure 2
\centering
\includegraphics[width=0.9 \linewidth]{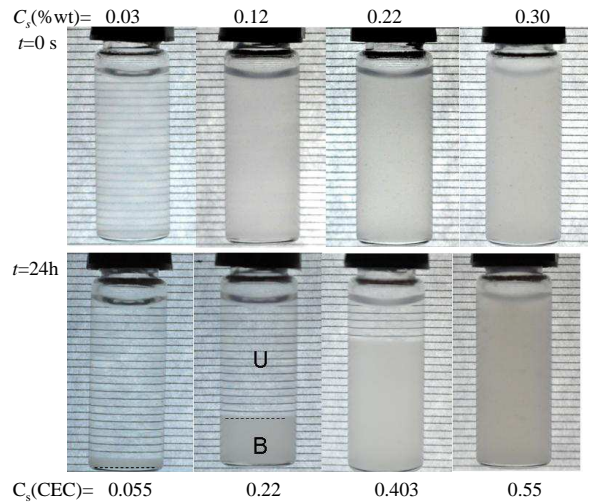}
\caption {\label{fig:Fig2}(Color online)
Photographs of fresh (a) and 24 h aged (b) suspensions of Laponite (2 \%wt).   Concentrations of CTAB, $C_s$, are shown in the figure. Letters U and B correspond to the upper and bottom layers, respectively.}
\end{figure}

Sedimentation stability in Earth gravity was checked by analyzing the photographs of fresh (a) and 24 h aged (b) suspensions with different concentrations of CTAB (Fig.\ref{fig:Fig2}). The fresh suspensions were visually homogeneous. However, even at small concentration of CTAB (above $0.03$ \%wt  $\approx 0.055 CEC$), they became turbid and unstable. The relatively fast separation of suspensions, few hours aged, was observed.  Finally, suspensions were separating out into the bottom (turbid) and upper (clear) phases, the volume of the bottom phase was continuously growing with $C_s$ increase and any suspension became turbid at $C_s \approx 0.3$ \%wt.

\begin{figure} [htbp]%Figure 3
\centering
\includegraphics[width=0.9 \linewidth]{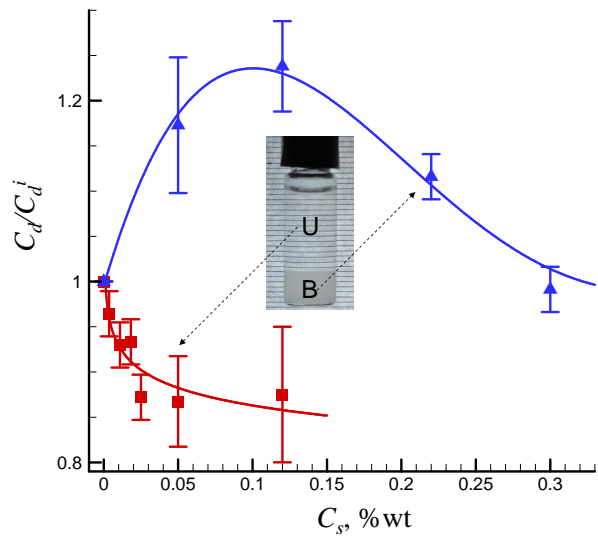}
\caption {\label{fig:Fig3}(Color online)
Ratio   of the concentrations of solids in the layers U(upper) and B(bottom), $C_d$, to the initial concentration of solids in the fresh suspension, $C_d^i$ , versus concentration of CTAB, $C_s$. Suspensions were 24 h aged.}
\end{figure}

Complementary analysis has shown that solid contents were different in the upper and bottom layers (U- and B-layers, respectively) (Fig. \ref{fig:Fig2}). The concentration of solids $C_d$ (i.e., the total concentration of Laponite and CTAB in water) was determined by separation of these layers and drying in desiccator for 24 h at 453 K. Figure \ref{fig:Fig3} presents  ratio versus concentration of CTAB, $C_s$, in U- and B-layers of 24 h aged suspensions. Here,  $C_d^i$  is the solid concentration in the fresh suspension.

The concentration of solids, $C_d$, was noticeably smaller in U- layer than in B- layer. However, concentrations of solids both in U- and B-layers were high. The observed separation of layers may reflect separation of phases with different contents of Laponite and CTAB. It is possible that less concentrated U-layer and more concentrated B-layer represent the phases of different non-ergodic states.

Note that our observations contradict to the phase behaviour of CTAB+Laponite suspensions described earlier \cite{Zhang2008}, where phase separation was not observed for small concentration of $C_s$( $<0.3 CEC$). However, it is not surprising, because another aging procedure was used in the reference \cite{Zhang2008}.  In order to clarify the possible effects of aging of the Laponite suspension, the dynamic light scattering (DLS) investigation were done at concentration of CTAB $C_s=0.0164$ \%wt. At such low CTAB concentration, the phase separation processes were insignificant, the bottom layer was very thin and the aging dynamics, probably, reflected the aging processes in the U-layer.

\begin{figure} [htbp]%Figure 4
\centering
\includegraphics[width=0.8 \linewidth]{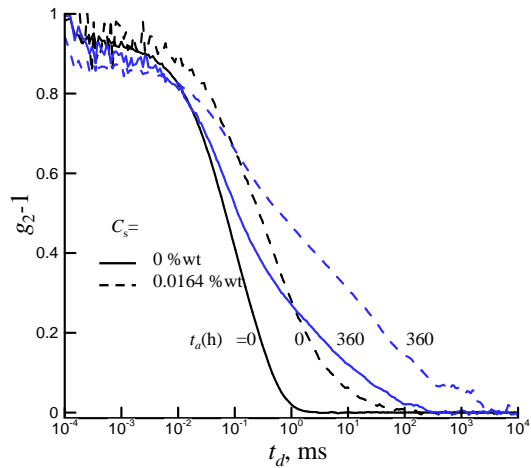}
\caption {\label{fig:Fig4}(Color online)
Examples of second order autocorrelation function, $g_2(t)-1$, versus decay time, $t_d$, for
freshly prepared and 360 h aged  suspensions in the absence (solid lines) and presence (dashed lines) of CTAB ($C_s=0.0164$ \%wt).}
\end{figure}

Figure \ref{fig:Fig4} presents examples of the second order autocorrelation function, $g_2-1$, versus the decay time, $t_d$, for freshly prepared and 360 h aged suspensions in the absence (solid lines) and presence (dashed lines) of CTAB. The introduction of CTAB evidently affected the shape of the autocorrelation function and aging dynamics. Supplementary analysis has shown that the shape of $g_2(t_d)$  may be well fitted using the following function
\begin{widetext}
\begin{equation}\label{Eq4}
 \sqrt{(g_2(t_d)-1)/b}=a\exp{(-t_d/\tau_1)}+(1-a)\exp {-(t_d/\tau_2)^ \beta},
\end{equation}
\end{widetext}
that was previously used to describe gelation process in the aqueous Laponite suspensions \cite{Abou2001, Cummins2007, Ferse2007, Jabbari-Farouji2008, Nicolai2000, Nicolai2001a, Rosta1990, Ruzicka2004a}.

This function represents the weighed sum of exponential and stretched exponential contributions that corresponds to the fast (single particles or small aggregates) and slow (the structural rearrangement of the system or large aggregates) motions. Here, the constant $b$ represents the coherence factor, $\tau_1$ and $\tau_2$ are the relaxation times of fast and slow motions, respectively, $a$ and $(1-a)$ are the amplitudes of these modes, and the stretching parameter $\beta(\leq 1)$ accounts for the polydispersity of aggregates. The value of $\beta=1$ corresponds to monosized aggregates. The mean relaxation time  $\tau_m$ was calculated as \cite{Ruzicka2004a}
\begin{equation}\label{Eq5}
\tau_m=\tau_2 \Gamma(1/\beta)/\beta,
\end{equation}
where $\Gamma$  is the usual Euler gamma function.

\begin{figure} [htbp]%Figure 5
\centering
\includegraphics[width=0.8 \linewidth]{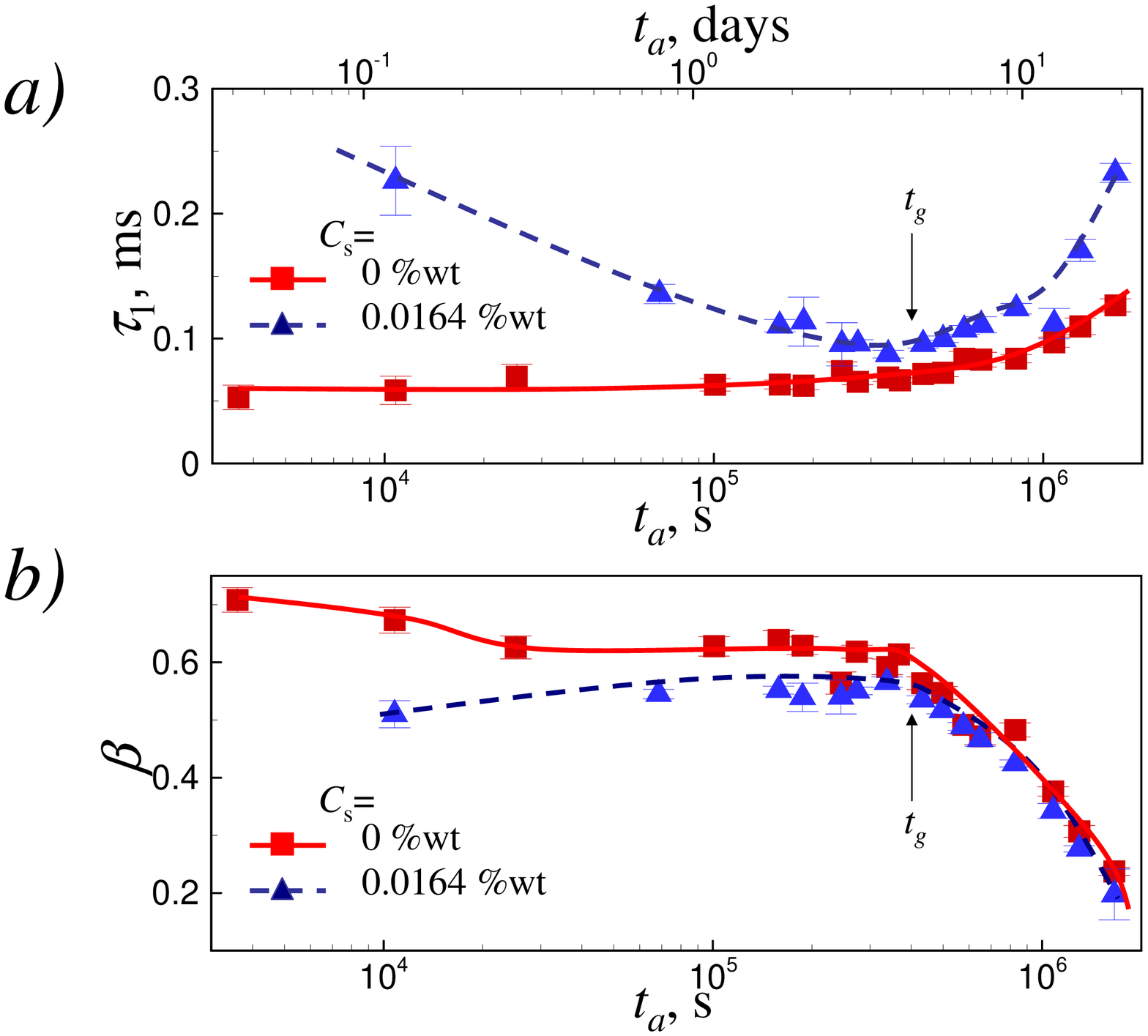}
\includegraphics[width=0.8 \linewidth]{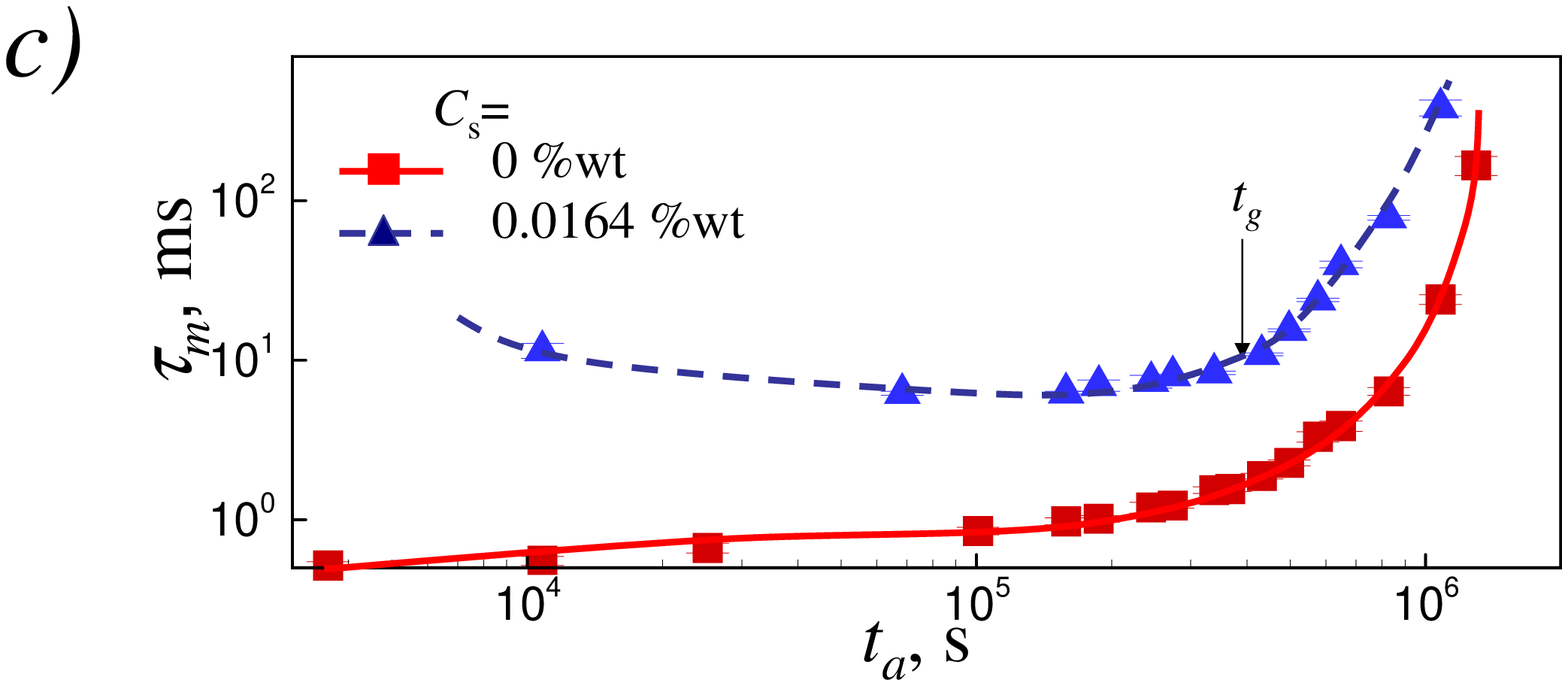}
\caption {\label{fig:Fig5}(Color online)
Effect of aging time $t_d$ on fast relaxation time $\tau1$(a), stretching parameter $\beta$ (b) and mean relaxation time $\tau_m$(c) in the absence (solid lines) and presence (dashed lines) of CTAB ($C_s=0.0164$ \%wt).}
\end{figure}

Figure \ref{fig:Fig5} presents the effect of the time of aging $t_a$ (0-460 h) on parameters $\tau_1$, $\beta$, and $\tau_m$.  In pure Laponite suspensions with Laponite concentration 2 \%wt, which corresponds to the boundary between phases IG$_1$ and IG$_2$ \cite{Ruzicka2011}, the aging resulted in increase of polydispersity, and both of relaxation time, $\tau_1$ and $\tau_m$. The most pronounced effects on  $\tau_1$, $\beta$ and $\tau_m$ were observed at $t_a>5^.10^5-10^6$ s (6-12 days) that was in correspondence with commencement of gelation, observed earlier for salt-free Laponite suspensions \cite {Ruzicka2004a}. From the other side, dependences  $\beta(t_a)$,  $\tau_1(t_a)$,  $\tau_m(t_a)$ demonstrated the presence of extremums at $t_a=t_g\approx 5^.10^5$ s ( 6 days) in the presence of CTAB. The character of these dependences at long time of aging ($t_a>t_g$) also can be explained by gelation processes. However, at early stages before gelation (i.e., at $t_a<<t_g$) the aging resulted in decrease of polidispersity and relaxation times $\tau_1$, $\tau_m$. The observed behaviour may be attributed to equilibration of CTAB molecules. Initial mixing of solutions of Laponite and CTAB may result in nonuniform distribution of CTAB molecules between different Laponite particles. During the process of equilibration in the course of the aging, the values of  $\beta$, $\tau_1$,  $\tau_m$  were approaching those representative for the CTAB-free suspensions (Fig.\ref{fig:Fig5}). It may reflect more homogeneous distribution of CTAB.
The data on  $m$ versus $t_a$ dependencies during gelation (i.e., at $t>t_g$) were fitted using the scaling equation \cite{Ruzicka2004a}
\begin{equation}\label{Eq6}
\tau = \tau _o\exp (\frac{B}{1-t_a/t_\infty}),
\end{equation}
and the following values were obtained for the most important fitting parameters: $B=7.0\pm 3.6$, $t =960\pm 360$ h  ($C_s=0$ \%wt) and $B=5.1\pm 1.4$, $t =640\pm 64$ h($C_s=0.0164$ \%wt).

The data obtained for pure 2 \%wt Laponite suspension in the absence of CTAB were in reasonable correspondence with data of reference \cite{Ruzicka2004a}($B\approx 6$, $t =700$ h) and evidenced that suspension was non-ergodic phase IG$_2$ at this concentration. Introduction of CTAB resulted in acceleration of gelation processes. However, the value of $B$ was still large and noticeably exceeded the values, characteristic for IG$_1$ phase ($B\approx 0.7$). So, it may be concluded, that introduction of CTAB at $0.0164$ \%wt not affected the state of repulsive gel IG$_2$, that  is inherent for the pure Laponite suspension at $C_l=2$ \%wt.

\subsection{\label{sec:AS}Accelerated stability analysis by analytical centrifugation}

In order to avoid aging processes, the accelerated stability analysis using the technique of analytical centrifugation was done.

\begin{figure} [htbp]%Figure 6
\centering
\includegraphics[width=0.8 \linewidth]{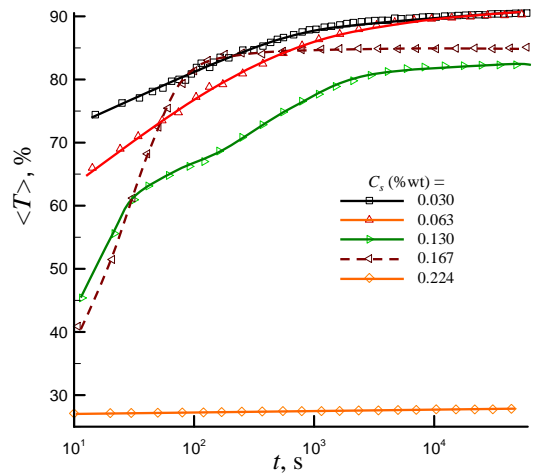}
\caption {\label{fig:Fig6}(Color online)
Mean light transmissionn, averaged over the height of the sample,  $<T>$ , versus time of centrifugation $t$ for different concentration of CTAB, $C_s$. The rotor speed $\omega$  was 500 rpm ($g=36.3g_o$).}
\end{figure}

Figure \ref{fig:Fig6} presents examples of mean light transmission  $<T>$  versus time of centrifugation $t$ for different concentrations of CTAB, $C_s$ and the same rotor speed  $\omega=500$ rpm ($g=36.3g_o$).
In such sedimentation at relatively high acceleration, the separation of bottom phase (similar to that observed in Fig. \ref{fig:Fig2}) was occurred practically instantly. Further time evolution of  $<T>$   reflected the changes in U- and B-layers. The transmission decreased (and turbidity increased) with increase of $C_s$ at low concentrations of CTAB ($C_s<0.13-0.15$ \%wt). In the course of centrifugation, the values of  $<T>$  smoothly increased and reached saturation at $t \geq10^4s$ ( 3 h). More accelerated time evolution of  $<T>$  was observed in the concentration interval between 0.14 \%wt  and 0.2 \%wt the (see, e.g., $C_s =0.167$ \%wt in Fig.      \ref{fig:Fig6}) and, finally, the value of  $<T>$  was practically constant in the course of sedimentation at $C_s$ above 0.2 \%wt ( $\approx 0.37 CEC$).
 	
\begin{figure} [htbp]%Figure 7
\centering
\includegraphics[width=0.8 \linewidth]{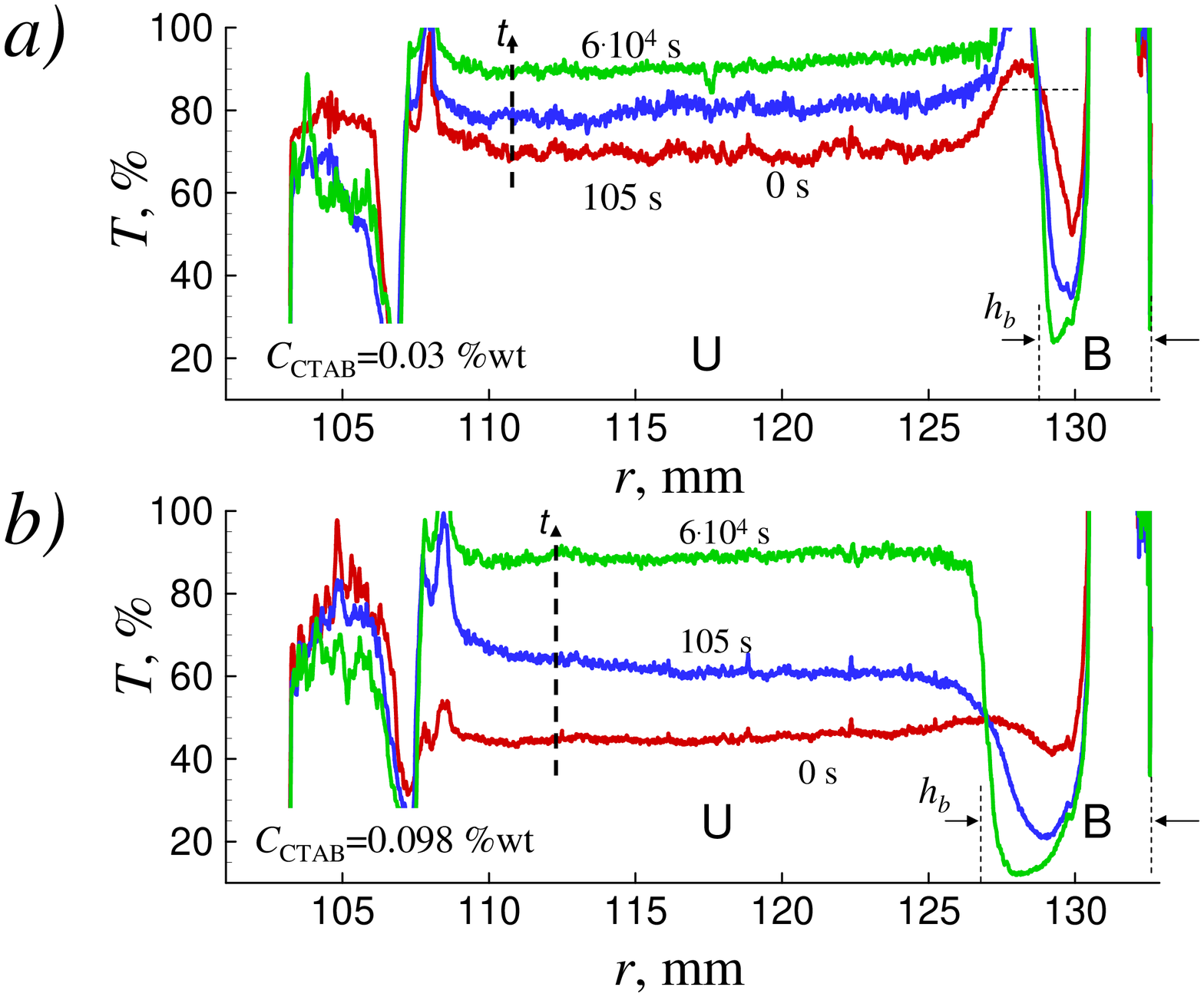}
\includegraphics[width=0.8 \linewidth]{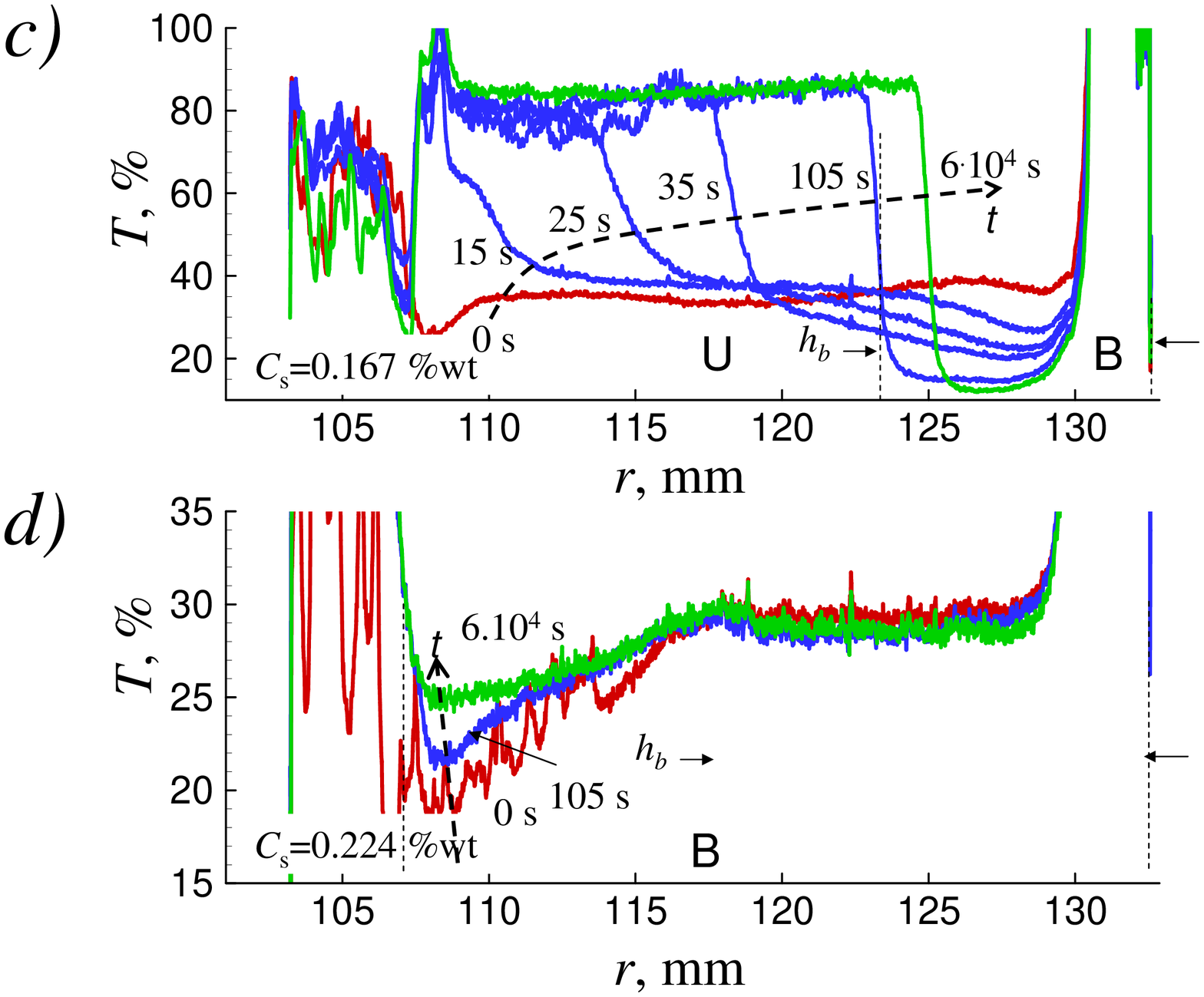}
\caption {\label{fig:Fig7}(Color online)
Profiles of light transmission $T(r)$ though the cell at different centrifugation times $t$ and different concentrations of CTAB $C_s=0.03$ \%wt (a), 0.098 \%wt(b), 0.167 \%wt(c), 0.224 \%wt(d). The rotor speed $\omega$  was 500 rpm ($g=32.3g_o$). Letters U and B correspond to the upper and bottom layers, respectively; $h_b$ is height of the bottom layer.}
\end{figure}

Figure \ref{fig:Fig7} presents the examples of light transmission profiles $T(r)$ in the course of sedimentation at different CTAB concentrations $C_s$=0.03 \%wt (a), 0.098 \%wt (b), 0.167 \%wt (c), 0.224 \%wt(d) and at the same rotor speed  $\omega=500$ rpm. At low concentrations of CTAB ($C_s<0.13-0.14$ \%wt), the height of B-layer $h_b$ was practically constant (Fig. \ref{fig:Fig7}a,b) during sedimentation. Transmissions of the U- and B-layers continuously increased and decreased, respectively. The transmission profiles $T(r)$ inside U-layer were approximately homogeneous at low concentrations of CTAB. For definiteness this sedimentation regime may be called as continuous-like (I).

However, the evolution of transmission profiles $T(r)$ became more complex at certain critical concentration ($C_s\approx 0.14$ \%wt). At the initial moment, the thickness of the B-layer phase $h_b$ reached the value comparable with the height of suspension $h$. The value of $h_b$ decreased and the bottom phase got compressed in the course of sedimentation, i.e. it was "soft" (Fig. \ref{fig:Fig7}c). The noticeable effects of zone-like (II) sedimentation were observed only at small time intervals ($t<200-300$ s). The level of transmission $T$ was approximately the same inside the U-layer and noticeably lowered inside the B-layer, i.e., the B-layer was becoming more and more dense. Finally, behaviour of the profiles $T(r)$ became stable at long time of sedimentation ($t>500$ s).

The "rigid" B-layer was formed at CTAB concentrations above $0.2$ \%wt ( $\approx0.37 CEC$). For example, at $C_s=0.224$ \%wt, the relative height 
%$h_b/h$( $\approx0.6$) 
and light transmission of the B-layer were practically constant in the course of sedimentation and only small changes in light transmission of the U-layer were observed (Fig. \ref{fig:Fig7}d). The "rigid" B-layer can be identified as a gel phase. For definiteness, this sedimentation may be called as gel-like (III) regime.
\begin{figure} [htbp]%Figure 8
\centering
\includegraphics[width=0.8 \linewidth]{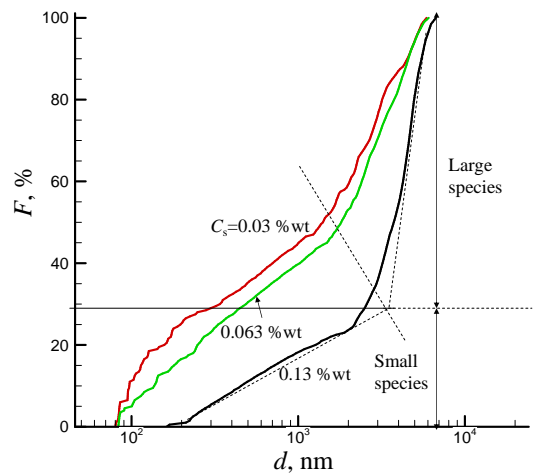}
\caption {\label{fig:Fig8}(Color online)
Volume-weighted cumulative distribution function $F(d)$ at different concentrations of CTAB, $C_s$.}
\end{figure}

Figure \ref{fig:Fig8} shows the volume-weighted cumulative distribution function $F(d)$, determined from U-layer sedimentation data at  $\omega=500$ rpm for CTAB concentration interval between 0.03 \%wt and 0.13 \%wt. The observed shape of $F(d)$ evidently reflected the presence of large and small species in CTAB+Laponite suspensions. The fraction of large species (they may correspond to the large aggregates of Laponite) continuously increased with increase of $C_s$. The critical concentration of transition to the zone-like sedimentation ($C_s \approx 0.14$ \%wt), possibly, reflects transformation of the CTAB+Laponite suspension to the state with dominance of large aggregates.

\begin{figure} [htbp]%Figure 9
\centering
\includegraphics[width=0.8 \linewidth]{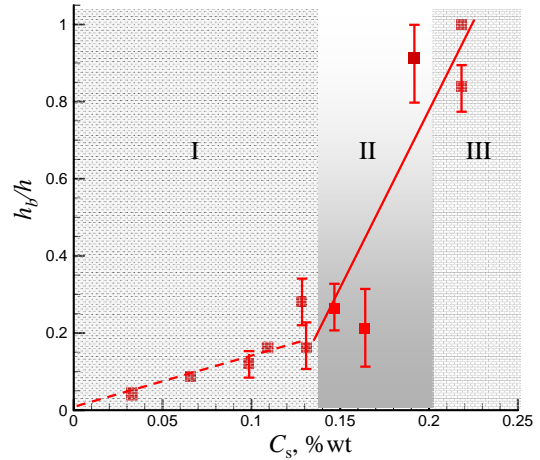}
\caption {\label{fig:Fig9}(Color online)
Relative height of the B-layer hb/h versus CTAB concentration $C_s$. The total time of centrifugation at the rotor speed  $\omega = 500$ rpm ($g=36.3g_o$) was  $t\approx 10^4$s. Different CTAB concentration ranges correspond to the regimes of continuous (I), zone-like (II) and gel-like (III) sedimentation.}
\end{figure}
Figure \ref{fig:Fig9} presents the relative height of the B-layer hb/h versus CTAB concentration $C_s$. The total time of centrifugation was  $10^4$ s and rotor speed was  $\omega=500$ rpm. The different CTAB concentration ranges correspond to the regimes of continuous (I), zone-like (II) and gel-like (III) sedimentation. It is interesting that the height of B-layer increased linearly with increase of $C_s$ within the range of continuous (I) sedimentation:
\begin{equation}\label{Eq7}
h_b/h=bC_s
\end{equation}
where $b=1.27 \pm 0.03$.

The supplementary ramping experiments have shown that the height of the "soft" B-layer in the regime of zone-like sedimentation (II) was sensible to the centrifugal acceleration $g\propto\omega^2$(see, Eq.\ref{Eq2}). In these experiments, the centrifugation was done with sequential increase of the rotor speed  $\omega=1000$ rpm ( $\approx 145 g_o$), 2000 rpm ( $\approx 581g_o$), 3000 rpm ( $\approx 1308g_o$), and 4000 rpm( $\approx 2326 g_o$). The centrifugation time  $\Delta t_\omega$  at the given $\omega$  was rather long and sufficient to obtain the stabilized light transmission profiles.

\begin{figure} [htbp]%Figure 10
\centering
\includegraphics[width=0.8 \linewidth]{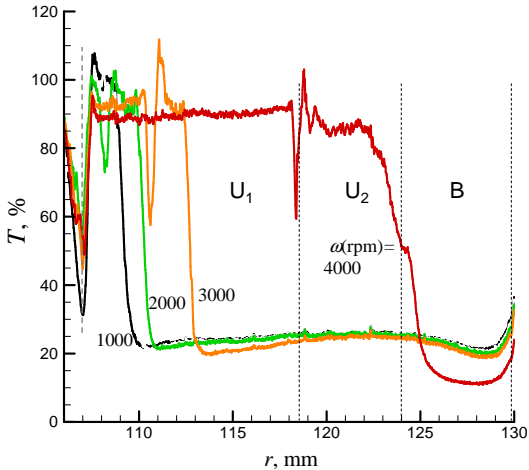}
\caption {\label{fig:Fig10}(Color online)
Profiles of light transmission $T(r)$ though the cell in the ramping centrifugation experiments with sequentially increased rotor speed  $\omega=1000$ rpm ( $t =6800$ s), 2000 rpm ( $t =14200$ s), 3000 rpm ( $t =7000$ s), and 4000 rpm ( $t =8000$ s). Here,  $t$  is the time of centrifugation at the given $\omega$. The concentration of CTAB was $C_s=0.19$ \%wt. The U-layer is divided by spike into U$_1$- and U$_2$-layers.}
\end{figure}

Figure \ref{fig:Fig10} shows the examples of light transmission profiles $T(r)$, obtained in these experiments for $C_s=0.19$ \%wt, which corresponds to regime zone-like (II) sedimentation. It is remarkable that increase of $\omega $  resulted in supplementary compressing of the B-layer. Moreover, light transmission profiles, obtained at  $ \omega=2000$ rpm, revealed the spike within the U-layer (Fig. \ref{fig:Fig9}). It evidently reflects the presence of two phases within the U-layer denoted as U$_1$ and U$_2$. The nature of these phases is still unclear and requires more thorough investigation in future.

\begin{figure} [htbp]%Figure 11
\centering
\includegraphics[width=0.83 \linewidth]{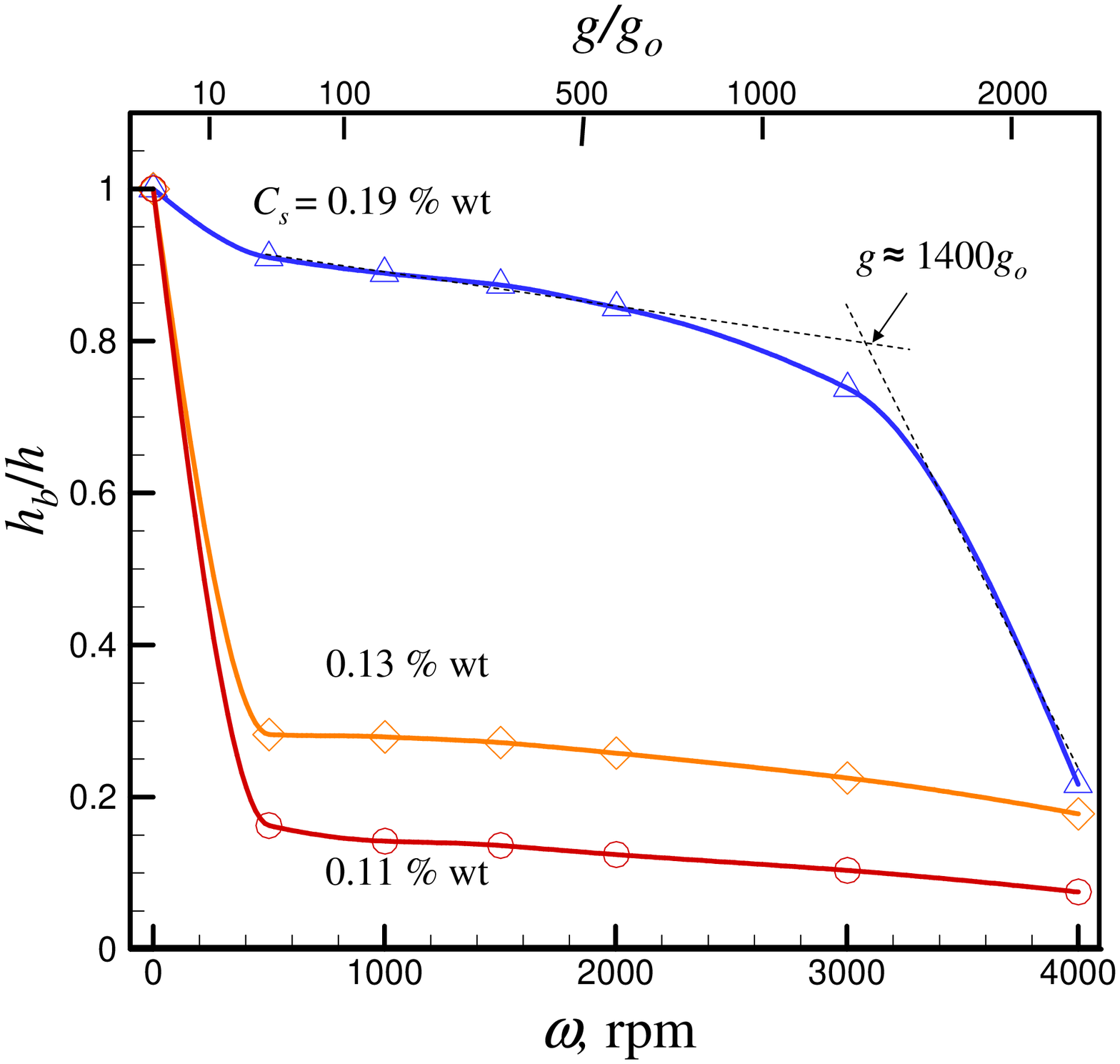}
\caption {\label{fig:Fig11}(Color online)
Relative height of the B-layer phase $h_b/h$ versus rotor speed $\omega$  at different concentrations of CTAB $C_s$. The time of centrifugation was fixed at 2h.}
\end{figure}

The differences between the regimes of continuous (I) and zone-like (II) were clearly manifested in dependence of the relative height of the B-layer $h_b/h$ versus rotor speed $\omega$  (Fig. \ref{fig:Fig11}). In these experiments, the time of centrifugation was constant, $t= 2$ hours. The value of $h_b/h$ decreased ( linearly) as value of $\omega$  increased for CTAB concentrations that correspond to the regime of continuous (I) sedimentation ($C_s\leq 0.14$ \%wt, Fig. \ref{fig:Fig9}). For higher concentrations of CTAB that correspond to the regime of zone-like (II) sedimentation (0.14 \%wt $\leq C_s \leq$ 0.20 \%wt), the collapse of the "soft" B-layer layer was observed at certain centrifugal acceleration. E.g., this collapse was observed at $g \approx 1400g_o$ for $C_s=0.19$ \%wt (Fig. \ref{fig:Fig11}). It corresponded to the critical damage of "soft" B-layer under the compression, caused by centrifugal acceleration.

\section{\label{sec:Conclusions}Conclusions}

This paper discusses sedimentation stability and aging of aqueous suspensions of Laponite in the presence of CTAB. The concentration of Laponite ($C_l=2$ \%wt) was corresponding to the boundary between IG$_1$ and IG$_2$ gels \cite{Ruzicka2011}. Both of these phases are stable against sedimentation and unstable against aging \cite{Joshi2008, Knaebel2000, Labanda2008}. It is expected that adsorption of CTAB on the surface of Laponite can increase its hydrophobicity and size of aggregates and decrease sedimentation stability of its suspensions. In Earth gravity, the sedimentation stability of suspension was violated even at small concentration of CTAB, suspension became turbid and separated into upper and bottom layers (U- and B-layers, respectively).  The dynamic light scattering (DLS) technique revealed that introduction of CTAB even in rather small concentration, $C_s=0.0164$ \%wt ($\approx 0.03 CEC$), induced noticeable changes in aging dynamics of U-layer. The most pronounced aging effects were observed after 6-12 days and they reflected commencement of gelation. At early stages before gelation, the anomalous behaviour of DLS was observed in the presence of CTAB. It may be attributed to equilibration of CTAB molecules nonuniformly distributed between different Laponite particles. The obtained data evidence that CTAB accelerates gelation. Accelerated stability analysis by means of analytical centrifugation revealed three sedimentation regimes:  continuous (I, $C_s<0.14$ \%wt), zone-like (II, $0.14 <C_s<0.2$ \%wt) and gel-like (III, $C_s >0.2$ \%wt). The B-layer was "rigid" in the I-st and III-rd regimes and "soft" in the II-nd -regime. In the regime of continuous sedimentation (I-st regime), the fraction of large aggregates in the U-layer and the height of B-layer increased with increase of $C_s$. The height of the "soft" B-layer in the regime of zone-like sedimentation was sensible to the value of rotor speed $\omega$. Increase of  $\omega$ resulted in supplementary compressing of B-layer and spitting of the upper layer into U$_1$ and U$_2$ layers. Moreover, the collapse of the "soft" B-layer at certain critical centrifugal acceleration was observed.

\begin{acknowledgments}
VS would like to acknowledge the support the Institute Charles Sadron, National Center of Scientific Research of France and Ministry of Education and Science of Ukraine (grand 014/60-SP). The authors also thank Dr. N.S. Pivovarova for her help with the manuscript preparation.
\end{acknowledgments}

%\bibliography{Manuscript}

\end{document}